\documentclass[nohyper,11pt,letterpaper]{JHEP3}
\usepackage{epsfig}

\newcommand{\beq}[1]{\begin{equation}\label{#1}}
\newcommand{\eeq}{\end{equation}}
\newcommand{\bear}[1]{\begin{eqnarray}\label{#1}}
\newcommand{\ear}{\end{eqnarray}}

\renewcommand{\theequation}{\arabic{section}.\arabic{equation}}
\catcode`\@=11 \@addtoreset{equation}{section}\catcode`\@=12
\newcommand{\np}{ {\newpage } }
\newcommand{\N}{ \mbox{\rm I$\!$N} }
\newcommand{\R}{ \mbox{\rm I$\!$R} }

\newcommand{\sign}{ \mbox{\rm sign} }
\newcommand{\e}{ \mbox{\rm e} }

\newcommand{\p}{\partial}

\newcommand{\btu}{\bigtriangleup}

\newcommand{\sums}{\sum\limits}
\newcommand{\const}{\mathop{\rm const}\nolimits}
\newcommand{\im}{{\rm i}}
\newcommand{\sh}{\ \rm sh}
\newcommand{\ch}{\ \rm ch}
\newcommand{\EFigure}[3]{\begin{figure}
        {\epsfxsize=#1mm\epsfbox{#2}}
        \caption{\protect\small #3}\end{figure}}
\author{V. D. Ivashchuk\\
    Center for Gravitation and Fundamental Metrology,\\
    VNIIMS, 3-1 M. Ulyanovoy Str., Moscow, 119313, Russia \\
    E-mail: \email{ivas@rgs.phys.msu.su}
}
\author{V. N. Melnikov\\
    Center for Gravitation and Fundamental Metrology,\\
    VNIIMS, 3-1 M. Ulyanovoy Str., Moscow, 119313, Russia \\
    E-mail: \email{melnikov@rgs.phys.msu.su}
}
\author{A.B. Selivanov\\
    Institute of Gravitation and Cosmology,\\
    Peoples' Friendship University of Russia, 6 Miklukho-Maklaya Str.,
    Moscow 117198, Russia \\
    E-mail: \email{seliv@rgs.phys.msu.su}
}

\abstract{ A family of cosmological solutions with  $(n+1)$
Ricci-flat spaces in the theory with several scalar fields and
multiple exponential potential is obtained when coupling vectors
in exponents obey certain relations. Two subclasses of solutions
with power-law and exponential behaviour of scale factors are
singled out.
 It is  proved that power-law solutions may take place
only when coupling vectors  are linearly independent and
exponential dependence  occurs for linearly dependent set of
coupling vectors. A subfamily of solutions with accelerated
expansion  is singled out.
 A generalized isotropization behaviours
of certain classes of general solutions are found.
In quantum case  exact solutions to Wheeler-DeWitt equation
are obtained and special "ground state" wave functions are
considered.
}

\title{Cosmological solutions \\
in multidimensional model with multiple exponential potential}

\begin{document}

\section{\bf Introduction}
\setcounter{equation}{0}

Recently,  the discovery of the cosmic acceleration
\cite{Riess,Perl} stimulated a lot of papers on multidimensional
cosmology with the aim to explain this phenomenon using certain
multidimensional models, e.g. those of superstring or supergravity
origin (see \cite{TownW}-\cite{BNQTZ}, and references therein).
These solutions usually deal with time-dependent scale factors of
internal spaces, thus overcoming a "no-go" theorem for static (and
compact) internal spaces \cite{Gibb1}. It should be noted that
certain part of publications does not deal with new exact
solutions but use old ones (sometimes rediscovered or written in
different parametrization).

A simple example of vacuum solution with acceleration was
considered by Townsend and Wohlfarth in \cite{TownW}. This is a
$(4+n)$-dimensional solution to vacuum Einstein equations with
$n$-dimensional internal space of negative curvature and
$4$-dimensional "our space" containing expanding $3$-dimensional
flat subspace. The solution from \cite{TownW} is a special
solution, in a  special model with a very special choice of
conformal frame. We note that it is a special case of more general
vacuum solution from \cite{I}, describing the "evolution" of
$(n-1)$ Ricci-flat spaces and one Einstein space of non-zero
curvature. The solution from \cite{I} was generalized to
scalar-vacuum case in \cite{BZ1,IM10,BZ2} and to the case of
composite p-brane configurations in \cite{IMJ} (for non-composite
case see also \cite{GrIM}). For a review see also \cite{IMtop}.
We  note that the solutions from \cite{I,IM10} may be also used
for generation of  special solutions
with several curved factor spaces (e.g.
from \cite{CHNOW}) using a curvature-splitting trick
\cite{GIM1} (when  Einstein space of non-zero
curvature is chosen as a product of several
Einstein spaces).
(For three integrable classes of
vacuum  cosmological solutions
with two factor spaces of non-zero curvature
in dimensions $D = 10, 11$ see \cite{GIM1}.)

At present rather popular models are those with  multiple
exponential potential of the scalar fields (see, for example,
\cite{RP,CopLW,BCopN,DGM,GPZ,GPCZ,NO} and refs. therein).

Such potentials also arise naturally in certain
supergravitational models \cite{Town}  and in sigma-models \cite{IMC},
related to configurations with $p$-branes.

Here we consider the $D$-dimensional model governed by the action

\bear{2.1i}
    S_{act} =  \int_{M} d^{D}z\sqrt{|g|} \{
    {\mathcal R}[g] - h_{\alpha\beta} g^{MN} \partial_{M}\varphi^\alpha\partial_{N}
    \varphi^\beta - 2 V_{\varphi}(\varphi) \} + S_{GH},
\ear
    with the scalar potential

\beq{pot}
    V_{\varphi}(\varphi) =
    \sum_{s \in S} \Lambda_s \exp[ 2 \lambda_s (\varphi) ] .
\eeq

Here $S_{\rm GH}$ is the standard Gibbons-Hawking boundary term
\cite{GH}.

\begin{list}{*}{The notations used here are the following ones:}
\item
    $\varphi=(\varphi^\alpha)$ is
    the vector from scalar fields in the space $\R^l$ with a metric
    determined by non-degenerate $l\times l$ matrix
    $(h_{\alpha\beta})$ with inverse one $(h^{\alpha\beta})$, \\
    $\alpha=1,\ldots,l$;
\item
     $\Lambda_s$ are constant terms, \\
     $s \in S$;
\item
    $\lambda_s$ is an $1$-form on $\R^l$:
    $\lambda_s (\varphi) =\lambda_{s \alpha} \varphi^\alpha$;\
    $\lambda^{\alpha}_{s} = h^{\alpha \beta} \lambda_{s \beta }$;
\item
    $g = g_{MN} dz^{M} \otimes dz^{N}$ is the metric,
    $|g| = |\det(g_{MN})|$, \\
    $M, N$  are world indicies \\
    (that may be numerated by $1, \ldots, D$);

\item
    $i, j = 0, \ldots, n$ are indicies describing
    a chain of factor spaces;\\
    $A = i, \alpha$ and $B = j, \beta$ are minisuperspace
    indicies \\
    (that may be numerated also by $0, \ldots, n+l$).

\end{list}

In this article we obtain new (and general)
families of classical and quantum ``cosmological''
solutions with vector coupling constants obeying

\beq{l1}
    \lambda_{s\alpha}\lambda_{s' \beta}h^{\alpha\beta}
      = \frac{D-1}{D-2}
\eeq
    for $s \neq s'$ and
\beq{l2}
    \lambda_{s\alpha}\lambda_{s \beta}h^{\alpha\beta}
    \neq \frac{D-1}{D-2}.
\eeq

    Here we find  exact  solutions with scale
factors and scalar fields depending upon one variable $u$ ("time"). We
keep the parameter $w = \pm 1$ in the metric $g= w
\e^{2{\gamma}(u)} du \otimes du + \ldots$ for a future
consideration of static configurations \cite{IMSp}. They may be of
interest in a context of "black hole" solutions with unusual
asymptotics (see \cite{Wilt,ChHMann,CaiJS,ChReall, BNQTZ} and
references therein).
We single out two new subclasses of solutions
with power-law and exponential behaviour of scale factors,
e.g. those with  accelerated  expansion.
A new result here is  generalized isotropization behaviours
of certain classes of classical  solutions that give
new examples of asymptotically accelerated expansion.

The paper is organized as follows. In Section 2 the model is
formulated. In Section 3 a class of classical cosmological
solutions corresponding to orthogonal $U$-vectors is presented.
Section 4 deals with special solutions exhibiting power-law and
exponential dependence of scale factors. Here we describe the
solutions with accelerated expansion. A special subsection is
devoted to a generalized isotropization. In section 5 quantum
analogues of the classical solutions  are considered. In Appendix
the equations of motion, corresponding to  (\ref{2.1i}) are
written (Appendix A) and derivations of special solutions of power
law and exponentional type are presented  (Appendix B and C); also
a classification of vectors in Euclidean space obeying relations
(\ref{l1}) and (\ref{l2}) is done (Appendix D).

We note that the classical and quantum
solutions for cosmological constant case
(i.e. for one term in potential with $\lambda_s =0$)
were considered earlier in \cite{IM} and \cite{BIMZ} for
vacuum and scalar-vacuum cases, respectively.

\section{\bf The model}

Let

\beq{2.10g}
    M = \R_{.}  \times M_{0} \times \ldots \times M_{n}
\eeq
    be a manifold equipped  with the metric

\beq{2.11g}
    g= w \e^{2{\gamma}(u)} du \otimes du +
    \sum_{i=0}^{n} \e^{2\phi^i(u)} g^i ,
\eeq
    where $w=\pm 1$, $\R_{.}$ is open interval in $\R$,
    $u$ is a distinguished coordinate;
    $g^i$ is a Ricci-flat metric on
    $d_{i}$-dimensional manifold $M_i$:

\beq{2.12g}
    {\mathcal R_{m_i n_i}}[g^i ] = 0,
\eeq
    $d_{i} = \dim M_i$,
$i=0,\dots,n$; $n\in {\bf N}$.

For dilatonic scalar fields we put
\beq{2.30n}
    \varphi^\alpha=\varphi^\alpha(u),
\eeq

It may be verified that the equations of motion corresponding to
(\ref{2.1i}) for the field configuration (\ref{2.11g}),
(\ref{2.30n}) are equivalent to equations of motion for
1-dimensional $\sigma$-model with the action

\beq{2.25gn}
    S_{\sigma} = \frac{1}2
    \int du {\cal N} \biggl\{G_{ij}\dot\phi^i\dot\phi^j
    +h_{\alpha\beta}\dot\varphi^{\alpha}\dot\varphi^{\beta}
    -2{\cal N}^{-2}V \biggr\},
\eeq
    where $\dot x\equiv dx/du$,
\beq{2.27gn}
    V =  -w V_{\varphi}(\varphi) \e^{2\gamma_0(\phi)}
\eeq
    is the potential ($V_{\varphi}$ is defined in (\ref{pot})) with
\beq{2.24gn}
    \gamma_0(\phi)
    \equiv \sum_{i=0}^{n}d_i\phi^i,
\eeq
    and
\beq{2.24gn1}
    {\cal N}=\exp(\gamma_0(\phi) -\gamma)>0
\eeq
    is lapse function. Here
\beq{2.c}
    G_{ij}=d_i\delta_{ij}-d_id_j, \qquad
    G^{ij}=\frac{\delta^{ij}}{d_i}+\frac1{2-D},
\eeq
    $i,j=0,\dots,n$,
 are components of  a gravitational part of minisupermetric and its dual
 \cite{IMZ}.

\subsection{Minisuperspace notations}

In what follows we consider minisuperspace $\R^{n+1 +l}$ with points
\beq{2.10}
    x \equiv(x^A)=(\phi^i,\varphi^\alpha)
\eeq
    equipped by  minisuperspace metric $\bar G = \bar G_{AB}dx^A\otimes
dx^B$ defined by the matrix and inverse one as follows:

\bear{2.35n}
    (\bar G_{AB})=\left(\begin{array}{cc}
    G_{ij}&0\\
    0&h_{\alpha\beta}
    \end{array}\right),\quad
    (\bar G^{AB})=\left(\begin{array}{cc}
    G^{ij}&0\\
    0&h^{\alpha\beta}
    \end{array}\right).
\ear

The potential (\ref{2.27gn}) reads

\beq{2.40n}
    V= - w  \sum_{s \in S} \Lambda_s\e^{2U^s(x)}
\eeq
    where $U^s(x)=U_A^sx^A$ are defined as follows

\beq{2.u}
    U^s = U^s(\phi,\varphi)= \lambda_{s\alpha}\varphi^{\alpha} +
    \sum_{i =0}^{n} d_i\phi^i,
\eeq
    or in components

\beq{2.38n}
    (U_A^s)=(d_i,\lambda_{s \alpha}).
\eeq

The integrability of the Lagrange system (\ref{2.25gn})
depends upon the scalar products of co-vectors $U^s$ corresponding
to $\bar G$:

\beq{2.45n}
    (U,U')=\bar G^{AB}U_AU'_B,
\eeq

    These products have the following form
\beq{2.48n}
    (U^s,U^{s'})= - b + \lambda_{s}\cdot\lambda_{s'},
\eeq
    where $s, s' \in S$ and

\bear{2.55n}
    \lambda_s\cdot\lambda_{s'} \equiv
    \lambda_{s\alpha}\lambda_{s' \beta}h^{\alpha\beta},\\
    b = \frac{D-1}{D-2}.
\ear

We put the orthogonality restriction on the vectors $U^s$

\beq{5.4n}
    (U^s,U^{s'})= -b + \lambda_s\cdot\lambda_{s'}=0,
\eeq
    $s\neq s'$, and also consider non-degenerate case
\beq{5.4na}
    (U^s,U^s) = -b + \lambda_s\cdot\lambda_s \ne 0
\eeq

(see (\ref{l1}) and (\ref{l2})).

In what follows we  denote
\bear{5.9n}
    h_s = (U^s,U^s)^{-1}\equiv \frac{1}{\lambda_s^2-b}.
\ear

The further consideration is based upon the
orthogonality conditions assumed.

\section{\bf Classical solutions}

Here we will integrate the Lagrange equations corresponding to the
action (\ref{2.25gn}) in the harmonic time gauge
$\gamma=\gamma_0$. We get a Lagrangian

\beq{5.31n}
    L=\frac12\bar G_{AB}\dot x^A\dot x^B-V,
\eeq
    where
$(\bar G_{AB})$ and $V$ are defined in (\ref{2.35n}) and
(\ref{2.40n}), respectively. The zero-energy constraint reads:

\beq{5.33n}
    E=\frac12\bar G_{AB}\dot x^A\dot x^B+V=0.
\eeq

The solutions to Lagrange equations corresponding to
 (\ref{5.31n}) in the
orthogonal and non-degenerate case (\ref{5.4n}), (\ref{5.4na})
read (see \cite{GIM,IMJ}):

\beq{5.34n}
    x^A =
    -\sum_{s\in S}\frac{U^{sA}}{(U^s,U^s)}\ln |f_s| + c^A u +
    \bar{c}^A,
\eeq

    Here
\bear{5.39n}
    f_s = R_s \sh(\sqrt{C_s}(u-u_s)), \;
    C_s>0, \; \epsilon_s<0; \\ \label{5.40n}
     R_s \sin(\sqrt{|C_s|}(u-u_s)), \;
    C_s<0, \; \epsilon_s<0; \\ \label{5.41n}
     R_s \ch(\sqrt{C_s}(u-u_s)), \;
    C_s>0, \; \epsilon_s>0; \\ \label{5.42n}
     |2 \Lambda_s/h_s|^{1/2}(u-u_s), \; C_s=0, \; \epsilon_s<0,
\ear
    where $u_s$ and $C_s$ are constants and

\bear{3.7a}
    R_s = |2 \Lambda_s/(h_s C_s)|^{1/2},\\
    \epsilon_s = - \sign(w \Lambda_s h_s).
\ear

Vectors $c=(c^A)$ and $\bar c=(\bar c^A)$ satisfy the linear
constraint relations

\bear{5.49n}
    U^s(c)= U^s_A c^A= \sum_{i =0}^{n} d_ic^i +
    \lambda_{s\alpha}c^\alpha=0,\\ \label{5.50n}
    U^s(\bar c)=  U^s_A \bar c^A= \sum_{i = 0}^{n} d_i\bar c^i +
    \lambda_{s \alpha}\bar c^\alpha=0.
\ear

The zero-energy constraint reads

\beq{5.53n}
    E=\sum_{s\in S} E_s+ \frac12 \bar G_{AB}c^Ac^B=0,
\eeq
    where  $C_s=2E_s(U^s,U^s)$, or, equivalently,
\beq{5.55n}
    \sum_{s\in S} C_s h_s +
    h_{\alpha\beta}c^\alpha c^\beta+\sum_{i=0}^n d_i(c^i)^2
    - \left(\sum_{i=0}^n d_ic^i\right)^2 = 0.
\eeq

{\bf The solutions.}

Using (\ref{5.34n}) and relations for contravariant components
$U^{sA} = \bar G^{AB} U^s_B$
\beq{4.8n}
    U^{si}= G^{ij}U_j^s= -\frac{1}{D-2}, \quad
    U^{s\alpha}= \lambda_s^\alpha,
\eeq
    we are led to relations for the metric

\bear{5.63n}
    g= \biggl(\prod_{s \in S} f_s^{2 b\ h_s}\biggr)
    \biggl\{ \e^{2c u+2\bar c} wdu\otimes du
    + \Bigl(\prod_{s\in S } f_s^{-2 h_s}\Bigr)
    \sum_{i = 0} \e^{2c^iu+2\bar c^i}g^i\biggr\}.
\ear
    and scalar fields

\beq{5.46n}
  \varphi^\alpha= - \sum_{s\in S}
   h_s \lambda_{s}^\alpha
  \ln |f_s|+c^\alpha u+\bar c^\alpha,
\eeq
$\alpha=1,\dots,l$.

Here
\beq{5.52aa}
  c = \sum_{i = 0}^{n} d_i  c^i, \qquad
  \bar c = \sum_{i = 0}^{n} d_i \bar c^i.
\eeq

The functions  $f_s$  are defined in (\ref{5.39n})--(\ref{5.42n})
and the relations on the parameters of solutions $c^A$, $\bar c^A$
$(A=i,\alpha)$ and $C_s$ imposed in (\ref{5.49n})--(\ref{5.50n}) and
(\ref{5.55n}), respectively. ($h_s$  are defined in (\ref{5.9n}).)

  \section{Special solutions}

Now we consider a special case of classical solutions
from the previous section when $C_s = u_s = c^i = c^{\alpha} = 0$
and
\beq{6.0}
   w \Lambda_s (\lambda_s^2 -b) > 0,
\eeq
$s \in S$.

We get two families of solutions written in synchronous-type
variable with:

  A) power-law dependence of scale factors for $B \neq -1$,

  B) exponential dependence of scale factors for $B  = -1$,

where

\beq{6.1}
  B = B(\lambda) = b\sum_{s \in S} h_s .
\eeq

Note that $h_s=h_s(\lambda_s)=(\lambda_s^2-b)^{-1}$.

\subsection{Power-law solutions}

Let us consider the solution corresponding to
the case $B   \neq -1$. The solution reads (see Appendix B)
\bear{6.2}
  g=  w d\tau \otimes d\tau
  + \tau^{2 \nu} \sum_{i = 0}^{n} A_i g^i, \\
  \label{6.3}
  \varphi^\alpha= - \frac{1}{B+1}\sum_{s\in S}
  h_s \lambda_{s}^{\alpha} \ln \tau
  + \varphi^\alpha_0,
\ear
where $\tau > 0$,

 \beq{6.4}
  \nu = \frac{B}{(B+1)(D-1)},
 \eeq
and

 \beq{6.5}
  |2 \Lambda_s/h_s| \exp( 2 \lambda_{s \alpha} \varphi^{\alpha}_0)
  = (B+1)^{-2},
 \eeq
$s \in S$; and $A_i > 0$ are arbitrary constants.

{\bf Solutions with "acceleration".}
Let us consider the
cosmological case $w = -1$. We get an accelerated
expansion of factor spaces if and only if $\nu > 1$
or, equivalently,
\beq{6.6}
 -b < B(\lambda) < -1.
\eeq

{\bf Remark 1. } For  $- 1 < B = B(\lambda) < 0$ we get $\nu < 0$, that
corresponds to a contraction of factor spaces. When
 $B < -b$ or $B > 0$ we  are led to the expansion with
deceleration, since $0 < \nu < 1$ in this case.
The case $B = 0$
corresponds to  static toy ``universe''. For $B = - b$
we obtain the expansion with zero acceleration ($\nu = 1$).

Let us consider the case when the matrix
$(h_{\alpha\beta})$ is positive
definite (e.g. $h_{\alpha\beta} = \delta_{\alpha\beta}$). It was
shown in Appendix D that the condition $B(\lambda) \neq -1$ in
this case implies that vectors $\lambda_s$ are linearly
independent. It follows also from Appendix D that the condition
$B(\lambda) < -1$ implies that

\beq{6.8}
    \lambda^2_{s_0} < b, \qquad
    \lambda^2_{s} > b \ {\rm for \ all \ }  s \neq s_0.
\eeq
Here $s_0 \in S$. These relations combined with inequalities
(\ref{6.0}) lead to
\beq{6.8a}
    \Lambda_{s_0} > 0, \qquad
    \Lambda_{s} < 0 \ {\rm for \ all \ } \  s \neq s_0,
\eeq
    i.e. one exponent term  in potential should be positive and
others should be negative.

    \subsection{Solutions with exponential scale factors}

Here we consider the solution corresponding to
the case $B = -1$. The solution reads (see Appendix C)
\bear{6.15}
    g=  w d\tau \otimes d\tau
    + \exp(2 m \tau) \sum_{i = 0}^{n} A_i g^i, \\
    \label{6.16}
    \varphi^\alpha=  (D-1) m \tau \sum_{s\in S}
    h_s\lambda_{s}^{\alpha}
    + \varphi^\alpha_0,
\ear
where
\beq{6.17}
    |2 \Lambda_s/h_s| \exp( 2 \lambda_{s \alpha} \varphi^{\alpha}_0)
    = m^2 (D-1)^2.
\eeq
$s \in S$; $m$ is parameter
 and $A_i > 0$ are arbitrary constants.

In the cosmological case $w = -1$ we get an accelerated
expansion of factor spaces if and only if $m > 0$.

Let us consider the case when $h_{\alpha\beta}$ is positive
definite. As it was shown in Appendix D the condition $B(\lambda)
= -1$ in this case  implies that vectors $\lambda_s$ are
linearly dependent, the linear term in (\ref{6.16}) vanishes and
hence

\beq{6.16a}
    \varphi^\alpha= \varphi^\alpha_0,
\eeq

i.e. for $(h_{\alpha\beta})$ of Euclidean signature all scalar
fields are constants when exponential expansion is under
consideration.
 We note that any subset of vectors $\lambda_s, s \in S \setminus \{ s_0
\}$ is linearly independent one in this case (see Appendix D).

{\bf Remark 2.} It may be shown that for the solution
(\ref{6.15})-(\ref{6.17}) the kinetic term for scalar fields
vanishes, i.e. $h_{\alpha\beta} g^{MN}
\partial_{M}\varphi^\alpha\partial_{N}\varphi^\beta = 0$ and the scalar
potential is constant: $ V_{\varphi}(\varphi) =
V_{\varphi}(\varphi_0) \equiv \Lambda$ and $(-w)2  \Lambda = m^2
(D -1)(D-2)$ (see (\ref{6.0})). For $w = -1$ and $\Lambda >0$ the
metric (\ref{6.15}) is coinciding with that for multidimensional
model from \cite{IM} with pure cosmological term $\Lambda$. Thus,
the exponential expansion of factor spaces is driven by effective
cosmological term $\Lambda$ and is not sensitive to time
dependence of scalar fields, that takes place only for
non-Euclidean signatures of   $(h_{\alpha\beta})$.

   \subsection{One-exponent example}

The simplest case with one term in the potential (\ref{pot}) is
interesting due to the absence of the restriction (\ref{5.4n}). It
follows from  eq. (\ref{6.0}) that the potential is positive:

\beq{6.7a}
    V_{\varphi} = \Lambda \e^{2
    \lambda_{\alpha}\varphi^{\alpha}},\quad
    \Lambda > 0
\eeq
    (we discard the indices in this example).

The accelerated power law  expansion takes place, if

\beq{6.7}
     0 < \lambda^2 < \frac{1}{D-2},
\eeq
    i.e. $\lambda^2$ should be
small enough but not equal to zero. The inequality is coinciding
(up to notations) with the eq. (1.9) from \cite{Town}. In this
case (\ref{6.4}) reads as $\nu = (\lambda^2 (D-2))^{-1}$.

For the exponential expansion eqs. (\ref{6.0}) and $B(\lambda) =
-1$ imply

\beq{6.7b}
    \lambda^2 = 0,
\eeq
    i.e. $\lambda$ should be zero
vector for positive definite matrix $(h_{\alpha\beta})$ or
light-like vector for  $(h_{\alpha\beta})$ of pseudo-Euclidean
signature.

    \subsection{Generalized isotropization}

In the case of one exponent the special solutions considered above
are attractors in the limit $\tau \to + \infty$ ($\tau$ is
"synchronous" variable) for general solutions with sinh-dependent
function $f_s$ (see (\ref{5.39n})), when  $\epsilon_s<0$, or,
equivalently, (\ref{6.0}) is satisfied and $B \leq -1$. This fact
may be readily verified using the relation $f_s \sim (u - u_s)$
for $u \to u_s$.

Here we consider a more general case of this attractor behavior
for multi-exponent case. Let $S_{-}$ be a subset of all $S$
satisfying $\epsilon_s<0$, or, equivalently, relations
(\ref{6.0}). Let us study a family of solutions with positive
$C_s$ and hence with $\sinh$-dependence of $f_s$ for $s \in S_{-}$
and $\cosh$-dependence of  $f_s$ for $s \in S \setminus S_{-}$.
Consider the solutions on the interval $(u_{*}, + \infty)$, where

\beq{max}
    u_{*}=  {\rm max}(u_s, s \in S_{-}).
\eeq

We denote by $S_{*}$ a subset of all $s \in S_{-}$  obeying $u_s =
u_{*}$. All functions $f_s(u - u_s)$ are smooth on this interval
and  near $u_{*}$  they behave as $f_s\sim (u -u_{*})$ for $s \in
S_{*}$, or as  (non-zero) constants for $s \in S \setminus S_{*}$.

Introducing the synchronous time variable by relation

\beq{time}
    \tau = \int du
    \exp(c u+\bar c) \prod_{s} (f_s(u - u_s))^{b h_s}
    + \const
\eeq
    we get $\tau \to +\infty$ as $u \to u_{*}$
in two cases:

A) $B_{*} < -1$ and
\beq{as1}
    \tau \sim (u - u_{*})^{B_{*} + 1};
\eeq

B) $B_{*} = -1$  and
\beq{as2}
    \tau \sim - \ln|u - u_{*}|.
\eeq

Here we denote
\beq{6.*}
    B_{*} = B(\lambda,S_{*}) = \sum_{s \in S_{*}} h_s b.
\eeq

{\bf  Case A.}
Using arguments analogous to those presented in Appendix B
we get the following asymptotical relations
in the limit $\tau \to +\infty$
for  $B_{*} < -1$
\bear{6.2A}
    g_{as}=  w d\tau \otimes d\tau
    + \sum_{i = 0}^{n} A_i\tau^{2 \nu_{*}} g^i, \\
    \label{6.3A}
    \varphi^\alpha_{as} = - \frac{1}{B_{*}+1}\sum_{s \in S_{*}}
    h_s \lambda_{s}^{\alpha} \ln\tau
    + \varphi^\alpha_0,
\ear
    where
\beq{6.4A}
  \nu_{*} = \frac{B_{*}}{(B_{*}+1)(D-1)},
\eeq
    and $A_i > 0$ are  constants.

Thus, we get an isotropization behaviour of  scale factors for
$\tau \to +\infty$, that is supported by exponential terms in the
potential labelled by $s \in S_{*}$. Other exponential terms do
not contribute to the power-law index $\nu_{*}$.

For
\beq{6.6A}
 -b < B_{*} < -1
\eeq
    we obtain the accelerated expansion for large enough values
of $\tau$.

{\bf  Case B.}

An analogous consideration in the case $B_{*} = -1$  leads
to the following asymptotical relations in the limit $\tau \to +\infty$
(see also Appendix C)
\bear{6.15B}
    g_{as}=  w d\tau \otimes d\tau +  \sum_{i = 0}^{n} A_i \exp(2
    m \tau) g^i, \\ \label{6.16B}
    \varphi^\alpha_{as}=  (D-1) m \tau \sum_{s
    \in S_{*}} h_s \lambda_{s}^{\alpha} +
    \varphi^\alpha_0,
\ear
where $m > 0$ and
\beq{6.17B}
    |2 \Lambda_s/h_s| \exp( 2 \lambda_{s \alpha} \varphi^{\alpha}_0)
    = m^2 (D-1)^2.
\eeq
$s \in S_{*}$.
Thus,  we obtain the accelerated expansion for large enough
values of $\tau$.

This is another regime of isotropization behaviour (for
one-exponent case with $\lambda_s = 0$ see also \cite{IM,BIMZ}).

{\bf Remark 3.} It may be shown that for the asymptotical solution
(\ref{6.15B})-(\ref{6.17B}) the kinetic term for scalar fields
asymptotically vanishes, i.e. $h_{\alpha\beta} g^{MN}
\partial_{M}\varphi^\alpha\partial_{N}\varphi^\beta \to  0$ as $\tau \to
+\infty$ and the scalar potential is asymptotically constant:

\beq{pot*}
    V_{\varphi}(\varphi) \to
    \sum_{s \in S_{*}} \Lambda_s \exp[ 2 \lambda_s (\varphi_0) ]
     \equiv \Lambda_{*}
\eeq
    as $\tau \to +\infty$. Here  $(-w) 2  \Lambda_{*} = m^2 (D
-1)(D-2)$. For $w = -1$ ans $\Lambda_{*} >0$ the ``asymptotical''
metric (\ref{6.15}) is coinciding with that for multidimensional
model from \cite{IM} with pure cosmological term $\Lambda_{*}$.
When $(h_{\alpha\beta})$ has Euclidean signature the exponential
isotropization behaviour takes place only for $S = S_{*}$ (it
follows from  Appendix D). In this case $\varphi_{as} =
\varphi_0$.

\subsection{Kasner-type behaviour}

Let us study the solution with $\sinh$-dependence in the limit
$u\rightarrow\infty$. We restrict ourselves by Kasner-like
behaviour of solutions for small $\tau$. The functions
(\ref{5.39n}) and (\ref{time}) behave as

\beq{kas.1}
    f_s \sim  \e^{\sqrt{C_s}u},\quad
    \tau \sim  \e^{(b \Sigma + c)u},\quad
    u\rightarrow + \infty,
\eeq
    where $\Sigma=\sum\limits_{s\in S} h_s \sqrt{C_s}$. To get the limit
$\tau\rightarrow + 0$ we put

\beq{kas.2}
    b\Sigma + c < 0.
\eeq

The metric and scalar fields in the limit $\tau\rightarrow +0$
read as follows:

\bear{kas.3}
    g_{as} = w d\tau\otimes d\tau + \sum_{i=0}^n \tilde{A}_i
    \tau^{2\alpha^i}g^i, \\ \label{kas.3a}
    \varphi_{as}^{\beta}=\alpha^{\beta}\ln\tau +
    \tilde{\varphi}_0^{\beta},
\ear
    where the Kasner parameters are given by the formulas

\bear{kas.4}
    \alpha^i = \frac{\Sigma\cdot (D-2)^{-1} + c^i}{b \Sigma + c},\\
    \alpha^{\beta} = \frac{- \sums_{s\in S} h_s \sqrt{C_s}\lambda_s^{\beta}
    + c^{\beta}}{b \Sigma + c}.
\ear

One can see that these parameters obey the generalized Kasner
relations:

\bear{kasner}
    \sum_{i=0}^n d_i \alpha^i = 1,\quad
    \sum_{i=0}^n d_i (\alpha^i)^2 +
    h_{\alpha\beta}\alpha^{\alpha}\alpha^{\beta} = 1.
\ear

We note that for $b\Sigma + c < 0$ we get a Kasner-like
asymptotics in the limit $\tau \rightarrow -\infty$ (with $\tau$
replaced by $|\tau|$ in (\ref{kas.3}) and (\ref{kas.3a})).

\subsubsection{Example: two spaces with one scalar field}

\EFigure{140}{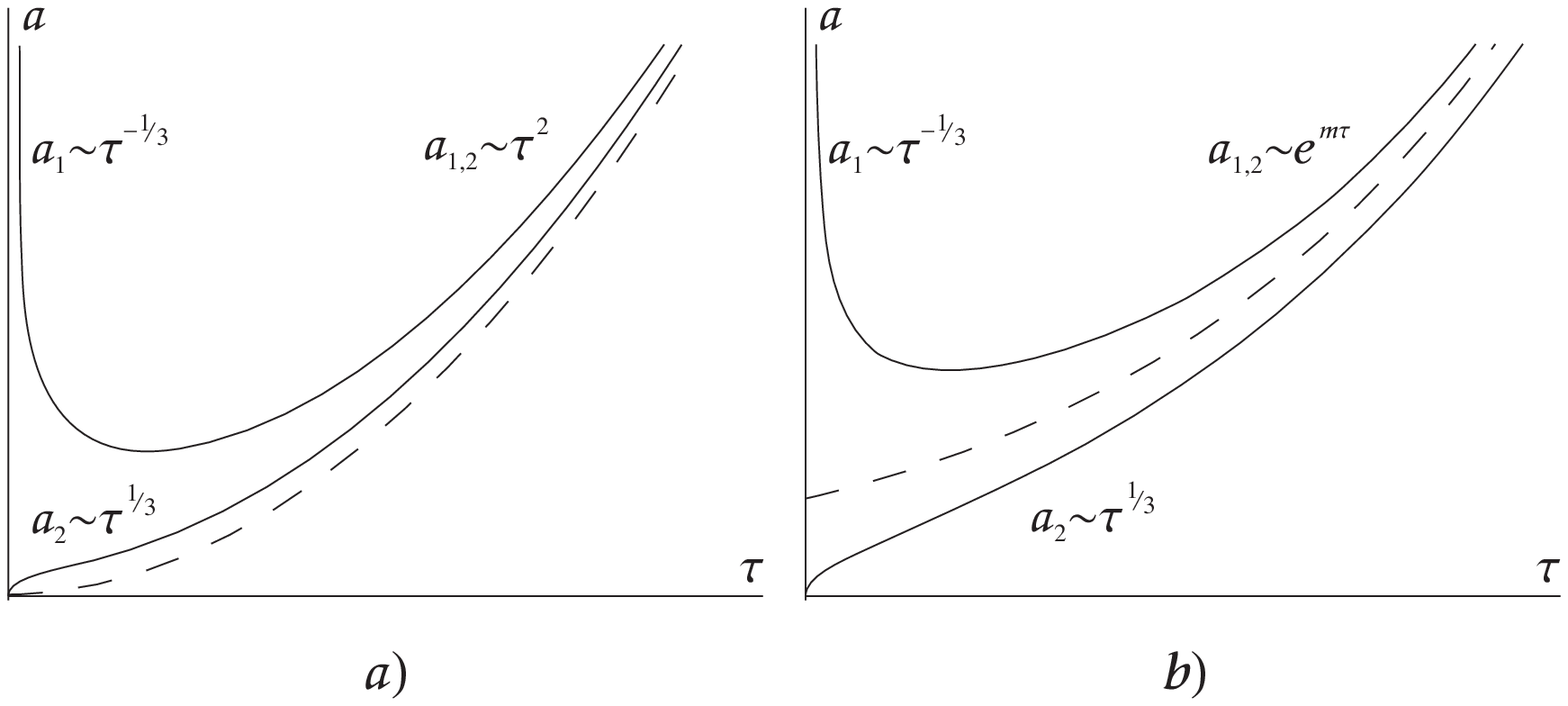} {The behaviour of scale factors of two
factor spaces of dimensions $d_0 =3$ and $d_1 =6$ in the presence
of one scalar field and one term in potential for the cases: a)
$\lambda^2 = 1/16$; b) $\lambda = 0$. In both cases the Kasner
parameter for the scalar field is chosen to be zero.}

As an example let us consider the model with two factor spaces,
one  scalar field and one term in potential.
It is not difficult to verify, that the parameters

\bear{kas.5}
    \alpha^0 = \frac{(d_0 - R)}{d_0(d_0 + d_1)},\quad
    \alpha^1 = \frac{(d_1 + R)}{d_1(d_0 + d_1)},\quad
    \alpha^{\beta}=0,
\ear

where $R = \sqrt{d_0 d_1 (D-2)}$, satisfy the eqs.
(\ref{kasner}).

For $d_1=3$, $d_2=6$ the behaviour of scale factors are presented
in Fig. 1, when: a) $\lambda^2 = 1/16$ (asymptotically power law
expansion); b) $\lambda = 0$ (asymptotically exponential
expansion).

{\bf Remark 4.} It will be shown in a separate publication
that Kasner-like behaviour near the singularity (as $\tau \rightarrow +0$)
is a generic one for $n > 0$ and a wide class of scalar potentials (\ref{pot}).

\section{Quantum solutions}

Here we find the quantum analogues of the classical solutions
presented above.

\subsection{General solutions}

The standard quantization of the energy constraint
$\mu E = 0$,  where $\mu \neq 0$ is a
parameter, leads to the Wheeler-DeWitt (WDW) equation
(see, for example, \cite{IMZ,Mis,Hal}) in the
harmonic-time gauge  $\gamma = \gamma_0(\phi)$
\beq{4.7n}
\hat H\Psi\equiv\left(-\frac{1}{2\mu}\Delta[{\bar G}]+\mu V\right)\Psi=0.
\eeq

Here $V$ is potential defined in (\ref{2.40n}), $\Delta[{\bar G}]$
is the Laplace operator, corresponding to ${\bar G}$, and $\Psi$
is wave function.

The  minisuperspace metric  may be diagonalized
by the linear transformation

\beq{5.12n}
    z^{\bar A}=S^{\bar A}{}_B x^B, \quad (z^{\bar A})=(z^0,z^a,z^s)
\eeq
    as follows
\beq{5.13n}
    \bar G_{AB}dx^A\otimes dx^B=-dz^0\otimes dz^0+
    \sum_{s\in S}\eta_sdz^s\otimes dz^s+dz^a\otimes dz^b\eta_{ab},
\eeq
    where $a,b=1,\dots,n+l-|S|$; $\eta_{ab} =\eta_{aa} \delta_{ab}$;
$\eta_{aa}= \pm 1$, $\eta_s=\sign (h_s)$, and

\beq{5.15n}
    q_sz^s=U^s(x),
\eeq
    with
\beq{5.17n}
    q_s = |h_s|^{-1/2} = \sqrt{| \lambda_{s}^2 - b|}>0,
\eeq
$s \in S$.

We are seeking the solution to WDW equation (\ref{4.7n}) by the method
of the separation of variables, i.e. we put
\beq{5.20n}
\Psi = \left(\prod_{s\in S}\Psi_s(z^s)\right)
\e^{\im p_az^a}.
\eeq

The wave function (\ref{5.20n}) satisfies to WDW equation
(\ref{4.7n}) if
\beq{5.22n}
    2\hat H_s\Psi_s\equiv\left\{-\eta_s \frac\partial{\partial z^s}
    \left(\frac\partial{\partial z^s}\right) -
    2 w \Lambda_s\e^{2q_sz^s}\right\}\Psi_s=2{\cal E}_s\Psi_s,
\eeq
    where

\beq{5.23n}
    \eta^{ab}p_ap_b+2\sum_{s\in S}{\cal E}_s=0.
\eeq
    (compare with classical relation (\ref{5.53n})).

The linearly independent solutions to eq. (\ref{5.22n})
reads
\beq{5.25n}
    \Psi_s(z^s)= B_{\omega_s}^s
    \left(\sqrt{-2 w \eta_s \Lambda_s}\, \frac{\e^{q_sz^s}}{q_s}\right),
\eeq
    where
\beq{5.27n}
    \omega_s= \sqrt{- 2{\cal E}_s h_s},
\eeq
    $s\in S$ and $B_\omega^s= a_s I_\omega + b_s K_\omega$
are superpositions of modified Bessel function.

The general solution to the WDW equation (\ref{4.7n}) is a superposition
of the "separated" solutions (\ref{5.20n}).

     \subsection{Special solutions}

Let us consider the special "ground state" solutions with $p_a =
{\cal E}_s = 0$ in (\ref{5.25n}), when relations (\ref{6.0}) are
imposed. These  solutions read

\beq{5.20m}
    \Psi = \prod_{s \in S}
    \left(a_s J_0(\sqrt{2 |h_s \Lambda_s |}\
    v \e^{\lambda_s(\varphi)})
    + b_s H^{(1)}_0
    (\sqrt{2 |h_s \Lambda_s|}\
    v \e^{\lambda_s(\varphi)})\right),
\eeq
    where $J_0$ and $H^{(1)}_0 $ are the Bessel and Hankel functions,
respectively, and $v = \exp(\sums_{i =0}^{n} d_i\phi^i)$ is a
volume scale factor. These quantum solutions correspond to special
classical solutions considered in the previous section.

We note that for large values of
\beq{5.20q}
       v_s = \sqrt{2|h_s\Lambda_s|}\ v \exp(\lambda_s(\varphi))
\eeq
    we get a quasi-classical regime, that may be obtained (along a
line as it was done in \cite{IKM}) using asymptotical relations
for Bessel functions. For small values of "quasivolumes" $v_s$ we
are led to a quantum regime. The crucial point here is that the
quantum domain is not defined only by volume scale factor $v$, but
by "quasivolumes" $v_s$, depending also upon scalar fields. In
strong enough scalar fields with certain signs (and the direction
of vector $\varphi$) and/or small $ |h_s \Lambda_s|$ one can
obtain a quantum behaviour of a wave function  for big enough
values of volume scale factor $v$. Analogous "effect" takes place
also for quantum solutions with "perfect fluid" \cite{IM10} when
certain equations of state are adopted.

\section{Conclusions}

Here we obtained a family of multidimensional cosmological
solutions with $(n+1)$ Ricci-flat spaces in the theory with
several scalar fields and multiple exponential potential.

The classical and quantum solutions are obtained if the
(orthogonality and non-degeneracy) relations (\ref{5.4n}) and
(\ref{5.4na}) on $U$-vectors are imposed, or, equivalently, when
relations (\ref{l1}) and (\ref{l2}) on coupling vectors are
satisfied. These solutions in fact correspond to $A_1 + ...+
A_1$ Toda-like solutions (for perfect fluid case see
\cite{GIM}).

Here we singled out the solutions with power-law and exponential
behaviours of scale factors. We proved that power-law dependence
may take place only when coupling vectors $\lambda_s$ are linearly
independent; exponential dependence  may occur only for a linearly
dependent set of  $\lambda_s$ obeying the condition $B(\lambda) =
-1$. We obtained the restriction on coupling vectors (\ref{6.6})
that cuts power-law solutions with acceleration. In subsection 4.3
the generalized isotropization behavior of certain class of
general solutions was found. Any asymptotics with power-law or
exponential behavior corresponds to a certain subset of
exponential terms in the potential (all other exponential terms do
not contribute to this asymptotical behaviour).

In the quantum case we solved the Wheeler-DeWitt equation and
singled out special "ground state" solutions that are quantum
analogous of special classical solutions from Section 4. These
solutions are defined as products of Bessel functions and depend
upon a set of quasivolumes that may be small enough (i.e.
belonging to "quantum domain") even when the volume scale factor
of the toy "universe" is in the classical region (in the Planck
scale).

\acknowledgments {

 This work was supported in part by the Russian Ministry of
 Science and Technology, Russian Foundation for Basic Research
 (RFFI-01-02-17312-a) and Project DFG (436 RUS 113/678/0-1(R)).

}

\renewcommand{\theequation}{\Alph{subsection}.\arabic{equation}}
\renewcommand{\thesection}{}
\renewcommand{\thesubsection}{\Alph{subsection}}
\setcounter{section}{0}

\section{Appendix}

\subsection{Equations of motion}

Here we outline for the sake of completeness the equations of motions
corresponding to the action (\ref{2.1i})
\bear{A.1}
    {\mathcal R}_{MN} - \frac{1}{2} g_{MN}{\mathcal R } =   T_{MN} ,
    \\    \label{A.2}
    {\btu}[g] \varphi^\alpha -
    \sum_{s \in S} 2 \lambda^{\alpha}_s
    e^{2 \lambda_s(\varphi)} \Lambda_s = 0.
\ear

In (\ref{A.1})
\bear{A.3}
    T_{MN} =
    h_{\alpha\beta}\left(\p_{M} \varphi^\alpha \p_{N} \varphi^\beta -
    \frac{1}{2} g_{MN} \p_{P} \varphi^\alpha \p^{P} \varphi^\beta\right)
    - V_{\varphi} g_{MN}.
\ear

\subsection{Power-law expansion}

Here we derive the solution (\ref{6.2})-(\ref{6.5})
from the general one. We start from the relation
(\ref{5.34n}) that can be written in our case as
follows
\beq{B.6}
    x^A(u)=
    -\sum_{s\in S}h_s U^{sA}\ln (\sqrt{|2 \Lambda_s/h_s|}u) +
    \bar{c}^A,
\eeq
where $u > 0$.

Introducing a new variable $\tau >0$ by formula
\beq{B.7}
u = C \tau^{1/(B+1)}, \qquad C > 0,
\eeq
we rewrite  (\ref{B.6}) in the following manner
\beq{B.8}
    x^A=
    -\sum_{s\in S}h_s U^{sA}\ln \tau + x^A_0
\eeq
with constants
\beq{B.9}
    x^A_0= \bar{c}^A - \sum_{s\in S}h_s U^{sA}
    \ln \left(\sqrt{|2 \Lambda_s/h_s|}C\right).
\eeq
    Due to orthogonality of $U^s$-vectors
the constraints on integration constant $U^s_A \bar{c}^A = 0$
may be written in the equivalent form
\beq{B.10}
    U^s_A x^A_0=  - \ln (\sqrt{|2 \Lambda_s/h_s|} C).
\eeq

In components the solution (\ref{B.8}) for
$(x^A) = (\phi^i, \varphi^{\alpha})$ reads
\bear{B.11}
    \phi^i=  \frac{B}{(B+1)(D-1)} \ln \tau + \phi^i_0, \\
    \label{B.12}
    \varphi^\alpha= - \frac{1}{B+1}\sum_{s\in S}
    h_s \lambda_{s}^{\alpha} \ln\tau
    + \varphi^\alpha_0.
\ear
Here $(x^A_0) = (\phi^i_0, \varphi^{\alpha}_0)$.

For $\gamma_0(\phi) = \sums_{i=0}^{n}d_i \phi^i$
we get
\beq{B.13}
\gamma_0(\phi) =
\frac{B}{(B+1)} \ln \tau + \phi_0,
\eeq
where
$\phi_0 = \sums_{i=0}^{n} d_i \phi^i_0$.

From the definition of "synchronous" variable
\beq{B.12.0}
 \exp(2\gamma_0(\phi) ) du^2 = d \tau^2
 \eeq
we obtain
\beq{B.13b}
C = |B+1| \exp( - \phi_0)
\eeq
and hence (\ref{B.10}) reads
\beq{B.14}
    \lambda_{s \alpha} \varphi^{\alpha}_0=  - \ln (\sqrt{|2 \Lambda_s/h_s|} C).
\eeq

The solution (\ref{6.2})-(\ref{6.5})
follows from the formulas (\ref{B.11}), (\ref{B.12}),
(\ref{B.12.0}), (\ref{B.14}) and $A_i =  \exp(2 \phi^i_0)$.

\subsection{Exponential expansion}

The solution (\ref{6.15})-(\ref{6.17})
may be obtained just along a line as it was done
for the power-law case.
The only difference here is the
relation
 \beq{C.18}
  u = C \exp(M \tau), \qquad C > 0
 \eeq
instead of (\ref{B.7}). Using a procedure analogous
to considered hereabove we get
\bear{C.19}
    \phi^i=  - \frac{1}{D-1}  M \tau + \phi^i_0, \\
    \label{C.20}
    \varphi^\alpha= - \sum_{s\in S}
    h_s \lambda_{s}^{\alpha} M \tau
    + \varphi^\alpha_0.
\ear
and
 \beq{C.21}
  \gamma_0(\phi) = - M \tau + \phi_0.
 \eeq
From (\ref{B.12.0})
we get
 \beq{C.19b}
   C = |M|^{-1} \exp( - \phi_0)
 \eeq
and introducing new parameter
 $m = - M/(D-1)$ we obtain the solution
(\ref{6.15})-(\ref{6.17}).

\subsection{$b-$proper set of vectors}

The statements of this subsection are based on relations
(\ref{l1}) and (\ref{l2})), that are used in Definition of
$b$-proper set (see below). The conditions for exponential
expansion follow from  Theorem.  Lemma 2 is useful for analysis of
"power-law" expansion.

In what follows $\{\lambda_i\}$ means a set of vectors $\lambda_1,
\ldots, \lambda_m \in \R^l, l \in \N$. The vector space $\R^l$ is
equipped with a scalar product (it corresponds to the positive
definite matrix $(h_{\alpha\beta})$ in the bulk of the article).

\textbf{Definition.} Let $b > 0$. The set of vectors
$\{\lambda_i\}$ is called  a $b-$proper one if

\bear{d.1}
    \lambda_i \cdot \lambda_j = b,\\
    \lambda_i^2\neq b,
\ear

$i\neq j, \ i,j=1,\ldots m.$

\textbf{Lemma 1.} Let $K=(\lambda_i \cdot \lambda_j)$ be a matrix
of scalar products for a $b-$proper set of vectors
$\{\lambda_i\}$. Then \beq{d.l1}
   {\rm det} K =
    (B+1)\prod_{i=1}^m (\lambda_i^2-b),
\eeq where

\beq{d.t0}
    B = B(\lambda_1, ..., \lambda_m)
    \equiv \sum_{i=1}^m \frac{b}{\lambda_i^2-b}.
\eeq

\Proof Using relations \beq{d.l2}
    K_{ij} = \lambda_i \cdot \lambda_j =
    (\lambda_i^2 - b)\delta_{ij} + b =
    (\lambda_i^2 - b)\left(\delta_{ij}+
    \frac{b}{\lambda_i^2 - b}  \right)
\eeq we represent the matrix $K$ as a product of a matrix $\bar K$
and a diagonal matrix $D$:

\bear{d.l3}
    \bar K_{ij}=\delta_{ij} + \frac{b}{\lambda_i^2 - b},\quad
    D_{ij}=\delta_{ij}(\lambda_i^2 - b).
\ear

A product of determinants ${\rm det} D$ and ${\rm det} \bar K$
yields (\ref{d.l1}). Here we used the relation ${\rm det} \bar K =
B+1$, that can be readily proved.

 \textbf{Theorem.} Let $\{\lambda_i\}$ be a $b-$proper set. The set
$\{ \lambda_i \}$ is linearly dependent if, and only if \beq{d.t1}
    B = B(\lambda_1, ..., \lambda_m) = -1.
\eeq In this case

\beq{d.t2}
    \sum_{i=1}^m \frac{\lambda_i}{\lambda_i^2-b} = 0
\eeq and any subset of $m-1$ vectors from $\{\lambda_i\}$ is
linearly independent one.

 \Proof The linear dependence of vectors
 $\{\lambda_i\}$ is equivalent to
 $\det K = 0$ that is equivalent (due to Lemma 1) to relation
(\ref{d.t1}).

From the relation
   \beq{d.t3}
    B(\lambda_1, ..., \lambda_m) =
    B(\lambda_1, ..., \lambda_{m-1}) + \frac{b}{\lambda_m^2-b}
    = -1
   \eeq
we get $B(\lambda_1, ..., \lambda_{m-1}) \neq -1$ and hence
the vectors $\lambda_1, ..., \lambda_{m-1}$ are linearly
independent. It is obvious that any $m -1$ $\lambda$-vectors are
also linearly independent.

To prove (\ref{d.t2}) let us denote

\beq{d.t4}
    \lambda=\sum_{i=1}^m \frac{\lambda_i}{\lambda_i^2-b}.
\eeq

Using the scalar products (\ref{d.l2}) we obtain

\beq{d.t5} \lambda^2= B(B + 1)/b = 0 \eeq
    and, hence, $\lambda=0$. The  Theorem is  proved.

As consequence we  obtain the inequality on the number of
$b$-proper vectors: $m \leq l+1$. (The equality $m = l+1$ takes
place when the vectors are linearly dependent.)

\textbf{Lemma 2.} In the $b-$proper set $\{\lambda_i\}$:

$a.$ there are no two or more vectors $\lambda_i$ obeying
$\lambda_i^2 < b$;

$b.$ $B > -1\ \Leftrightarrow$  vectors $\lambda_i$ are linearly
independent and all $\lambda_i^2 > b$.

\Proof $a.$ Let us suppose that there are two vectors $\lambda_i,\
 \lambda_j$, such that $\lambda_i^2 < b$ and $\lambda_j^2 < b$.
 Then $b^2 = (\lambda_i\cdot\lambda_j)^2 \leq
 \lambda_i^2\lambda_j^2 < b^2$, i.e. we come to a contradiction.

 $b.$ If $B>-1$ then by the Theorem the vectors are linearly
 independent and hence the matrix
 $(\lambda_i\cdot\lambda_j)=(K_{ij})$ is positive definite. This
 implies ${\rm det} K > 0$. Due to relation (\ref{d.l1}) and part
 $a.$ of this lemma we get: $\lambda_i^2 > b$ for all vectors.

Now, let the vectors $\{\lambda_i\}$ be linearly independent and
all $\lambda_i^2>b$. Then $\det K > 0$, and due to (\ref{d.l1}),
$B+1 > 0$.  The lemma is proved.

\np

\end{document}